\documentclass[twoside,fleqn]{article}
\usepackage{espcrc2}
\usepackage{epsfig}

\def\lsim{\raise0.3ex\hbox{$<$\kern-0.75em\raise-1.1ex\hbox{$\sim$}}}
\def\Chi{\raise0.4ex\hbox{$\chi$}}

\newcommand{\beqn} {\begin{equation}}
\newcommand{\eqn} {\end{equation}}
\newcommand{\refe}[1]{(\ref{#1})}

\title{Thermodynamics using Wilson and Staggered Quarks}

\author{E. Laermann\address{Fakult\"at f\"ur Physik,
        Universit\"at Bielefeld, Postfach 100 131, 33501 Bielefeld,
        Germany}%
        }
       
\begin{document}

\begin{abstract}
Recent developements in QCD at finite temperature are reviewed.
Particular emphasis is laid on results stemming from simulations
which involve quarks.
\end{abstract}

\maketitle

\section{INTRODUCTION}

Lattice investigations of QCD at finite temperature have
contributed considerably to the current understanding of
the transition from the hadronic state of matter to 
the quark gluon plasma and of the physics of the plasma
phase. 
Mainly due to the ever-present limitations of computational power
many analyses have been carried out in the pure gauge 
sector of QCD such that bulk properties of gluons at finite temperature
can be regarded as solved: the system has a well-established
first-order transition \cite{q_trans}, the equation of state
is known in the continuum limit \cite{BoydEOS} and the critical
temperature in the continuum limit has been determined with
only a few percent uncertainty \cite{Iwastring,Beinlich}.
Clearly, more detailed questions like e.g. the nature of
excitations in the plasma deserve further work,
also in the quenched approximation.
Yet, the emphasis of recent research has shifted towards
studies of full QCD including staggered as well as Wilson
quarks. 
These studies consistently have lead to an estimate of
the critical temperature of order 150 MeV for
2 flavors so far. This value is considerably lower than 
the quenched number of
$T_c = 270 (5) {\rm MeV}$. Through the relation
$T = 1/(a N_{\tau})$ 
of the temperature $T$ to lattice spacing $a$ and temporal extent
$N_{\tau}$ of the lattice,
dynamical fermion simulations in the vicinity of the transition
are, at a given $N_{\tau}$, carried out at considerably larger
lattice spacings. Already for that reason, 
with the standard discretizations,
extrapolations to the contiunuum limit will be more difficult
than in the quenched case. Therefore, the search for improved
actions has received much attention recently in the context
of finite temperature QCD.

This review attempts to summarize the developements in
finite temperature lattice QCD since
last year's conference as reviewed in \cite{Akira96}. 
I will concentrate
on problems and results in simulations with full QCD.
Section 2 summarizes estimates of the critical temperature.
In section 3 the present status of knowledge about the
nature of the chiral transition is discussed, mainly for 
two flavors staggered as well as Wilson fermions.
Section 4 describes studies of energy density and pressure
in the high temperature limit while section 5 reviews
a few recent results on screening lengths and masses.
This year's results at finite density are summarized
in section 6, conclusions are given in section 7.

\section{CRITICAL TEMPERATURE}

\begin{figure}[htb]
 \epsfig{file=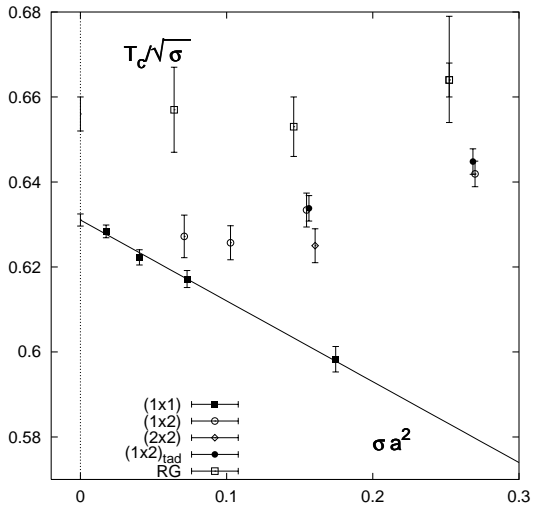,width=100mm}
\vspace{-15mm}
\caption{The quenched critical temperature in units of the square root
of the string tension for various gauge actions versus the 
the lattice spacing squared.}
\label{fig:tcquenched}
\end{figure}

One of the basic quantities to be derived from finite
temperature lattice QCD is the value of the critical
temperature. In order to prepare the stage for the
subsequent discussion of results from dynamical fermion
simulations, 
Figure~\ref{fig:tcquenched} summarizes the current status
of analysis in the quenched approximation \cite{Beinlich}. 
Figure~\ref{fig:tcquenched}
shows the ratio $T_c/\sqrt{\sigma}$ where $\sigma$ is the
string tension extracted from the static quark potential
for various actions. For all data points the value of the
critical coupling has been extrapolated to its infinite
(spatial) volume limit at which then the string tension
was determined. The lowest set of data points originates
from simulations with the standard Wilson gauge action
\cite{BoydEOS,Beinlich}.
A quadratic extrapolation in the lattice spacing to the
continuum limit gives $T_c/\sqrt{\sigma} = 0.631(2)$.
Symanzik-improved actions show a much weaker cut-off
dependence and are in agreement with the continuum
extrapolation of the standard action \cite{Beinlich,Cella}
Likewise,
the results \cite{Iwastring} from Iwasaki's RG-improved action 
are consistent
with a constant behavior in $a$, but they deliver a value
of the critical temperature of $T_c/\sqrt{\sigma} = 0.656(4)$
which is about 3\% higher than the number from the standard 
action. Since the procedure to extract the string tension
has not been the same for the two numbers, one might
suspect that the difference in the quoted value for $T_c$,
$T_c = 276(2) {\rm MeV}$ versus 266(2) MeV, is mainly due
to differences in the analysis of the static quark potential
\cite{Lego} rather than to differences in the improvement scheme.

\begin{figure}[htb]
 \epsfig{file=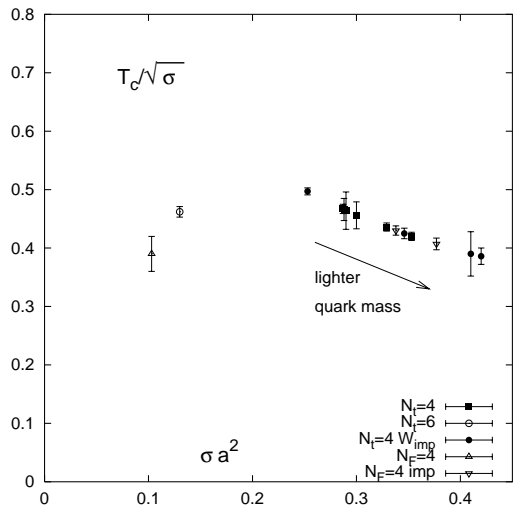,width=100mm}
\vspace{-15mm}
\caption{The critical temperature in units of the square root
of the string tension for dynamical fermions versus the 
square of the lattice spacing (for further explanations
see text).}
\label{fig:tcsigma}
\end{figure}

The current situation with dynamical fermions is depicted in 
Figure~\ref{fig:tcsigma}. The plot summarizes data from
simulations with 2 flavors of quarks, staggered fermions at
$N_{\tau} = 4$ and 6 \cite{MILC9703,oursigma} as well as
improved Wilson fermions at $N_{\tau} = 4$ \cite{MILC9703},
plus $N_F = 4$ staggered results obtained from $N_{\tau}=4$
lattices with an improved action \cite{impferm} in addition to an old
number \cite{MTc} from $N_{\tau} =8$ and a standard action.
Compared to the equivalent quenched plot, Figure~\ref{fig:tcsigma}
shows that the lattice spacings at which $T_c$ has been determined
so far 
are considerably larger than in pure gauge theory simulations.
Moreover, the investigations have not been carried out at the physical
quark masses. The arrow in Figure~\ref{fig:tcsigma} indicates
that at fixed $N_{\tau}$ the transition takes place at larger
lattice spacings when the quark mass is decreased. Thus
the critical temperature
is decreasing when the quark mass is lowered.

\begin{figure}[t]
 \epsfig{file=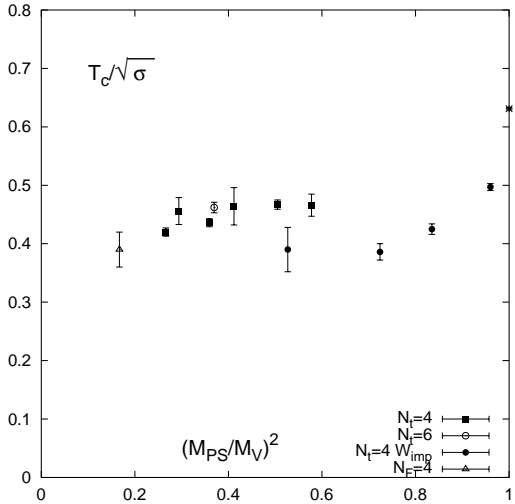,width=100mm}
\vspace{-15mm}
\caption{The critical temperature in units of the square root
of the string tension for dynamical fermions plotted versus 
$(M_{PS}/M_V)^2$. The point at $(M_{PS}/M_V)^2=1$ is the
$N_{\tau}=4$ quenched value.}
\label{fig:tcsigrho}
\end{figure}

The same data is shown again as a function of
the pseudoscalar Goldstone boson to vector meson mass ratio
$(M_{PS}/M_V)^2$ in Figure~\ref{fig:tcsigrho}.
(The meson masses at the appropriate critical values of 
the coupling in parts have been obtained by means of
a phenomenological interpolation formula \cite{MILCspec}.)
Here now, at fixed $N_{\tau}$, smaller lattice spacings are 
to the right of the figure.
The $N_{\tau} = 4$ staggered data indicates 
that $T_c$ over $\sqrt{\sigma}$ tends
to lower values as the quark mass is decreased. The same
trend is observed for the Wilson improved results, although
at larger $(M_{PS}/M_V)^2$ ratio. On the other hand, the
$N_{\tau} = 6$ data point seems to indicate that, at a given quark
mass, decreasing the lattice spacing increases $T_c/\sqrt{\sigma}$
only slightly. At the moment, one would therefore estimate
a physical value for the critical temperature of
$T_c/\sqrt{\sigma} \lsim 0.4$ or $T_c \lsim 170 {\rm MeV}$.

\begin{figure}[htb]
 \epsfig{file=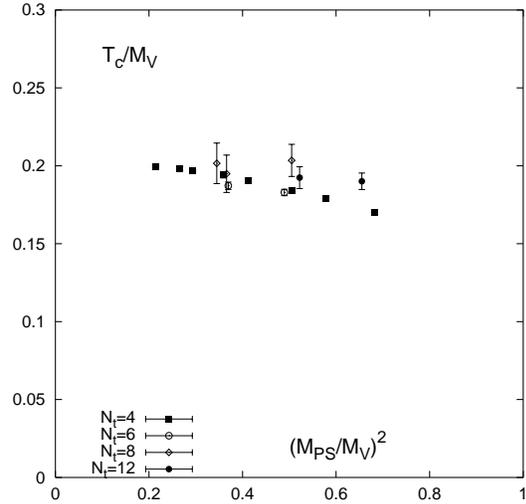,width=100mm}
\vspace{-15mm}
\caption{The critical temperature in units of the vector meson mass
for staggered fermions plotted versus 
$(M_{PS}/M_V)^2$.}
\label{fig:tcrhostagg}
\end{figure}

The critical temperature has also been estimated from
the ratio to the vector meson mass. In this case one ought
to go (close) to the chiral limit in order to extract
a physical number. In the case of using the string tension
to set the scale one might argue that the string tension
is considerably less affected by the quark mass. 
Figure~\ref{fig:tcrhostagg} shows $T_c/M_V$ for $N_F=2$
staggered fermions 
\cite{oursigma,Gottlieb,Bernard,HTMCGC,Bob,MILCspec}, 
plotted as function of $(M_{PS}/M_V)^2$.
As the quark mass is decreased this ratio rises. Recall that
the infinite quark mass, quenched data point corresponds
to $M_{PS}/M_V = 1$ and $T_c/M_V = 0$. 
As the lattice spacing is decreased, $T_c/M_V$ stays remarkably
constant. Extrapolating the $N_{\tau}=4$ data to the chiral limit
suggests a value of $T_c/M_V \simeq 0.2$ or $T_c \simeq 150 {\rm MeV}$.
Note that this value disagrees somewhat with the number
extracted from the string tension.

\begin{figure}[htb]
 \epsfig{file=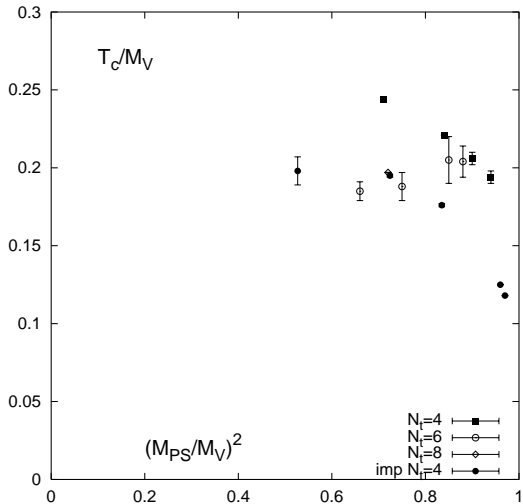,width=100mm}
\vspace{-15mm}
\caption{The critical temperature in units of the vector meson mass
for Wilson fermions versus $(M_{PS}/M_V)^2$.}
\label{fig:tcrhowilson}
\end{figure}

The corresponding data for dynamical Wilson quarks
\cite{MILC9703,Bitar,Bernard2,Bernard3} are given
in Figure~\ref{fig:tcrhowilson}. Despite the known problems
with standard Wilson fermion thermodynamics, at least the 
results for $N_{\tau} = 6$ and 8 as well as the first data
with improved Wilson fermions \cite{MILC9703} are not in disagreement with
the staggered data.

\section{PHASE TRANSITION}

\subsection{Staggered $N_F = 2$: critical behavior}

The theoretical expectations on the scaling behavior
of the theory at the chiral transition are based on the
$\sigma$ model in three dimensions.
For the case of two light flavors,
if the transition is second order, it is
expected to show scaling behavior with
$SU(2) \times SU(2) \simeq O(4)$
exponents. On the other hand, if the
anomalous $U_A(1)$ symmetry were effectively
restored, the relevant symmetry group would be
$U_A(1) \times SU(2) \times SU(2) \simeq
O(2) \times O(4)$ and the transition could be
first order \cite{WiPi}.

It has been attempted to analyze the critical behavior
of 2 flavor staggered QCD by studying the scaling
behavior of various quantities and determining
critical exponents \cite{FKEL}.
These scaling relations are
derived from the scaling of the singular part
of the free energy density,
\beqn
f(t,h) = -\frac{T}{V} \ln Z = b^{-1} f( b^{y_t}t, b^{y_h} h)
\label{free_energy}
\eqn
Here $T$ and $V$ denote temperature and volume, $Z$ is
the partition function, $t$ is the reduced temperature,
$t = (T-T_c)/T_c$, and $h$ is the symmetry breaking field,
$h = m/T$ where $m$ is the quark mass.
The scaling factor $b$
is arbitrary and can be chosen appropriately.
In the vicinity of the critical point thermodynamic
quantities should be governed by the thermal ($y_t$)
and the magnetic ($y_h$) critical exponent. In the
staggered version of lattice regularized QCD, for the
dimensionless couplings $t$ and $h$ one uses 
\begin{eqnarray}
t & = & \frac{6}{g^2} - \frac{6}{g_c^2(0)} \nonumber \\
h & = & m a N_{\tau}
\label{eq:def}
\end{eqnarray}
where $g_c(0)$ denotes the critical coupling
on a lattice with fixed temporal extent in the
limit of vanishing quark mass. At non-vanishing quark
mass, a pseudo-critical coupling $g_c(m)$ is defined
as the location of a peak in e.g. the Polyakov loop
susceptibility.

The status of investigations of the
chiral transition with two light staggered quarks has been commented
upon at last year's conference \cite{Akira96}. Since then, 
JLQCD \cite{JLQCDcrit} as well as the Bielefeld group \cite{us_crit}
have about finalized their analyses.
A new analysis was presented by C. DeTar at this conference
\cite{Carleton}.

The quantities analyzed by JLQCD and the Bielefeld group are 
various susceptibilities, in particular
the magnetic or chiral susceptibility
\beqn
\chi_m = \frac{T}{V} \sum_{i=1}^{N_F}
        \frac{\partial^2}{\partial m_i^2} \ln Z
\label{suscep_mag}
\eqn
and the thermal susceptibility
\beqn
\chi_t = -\frac{T}{V} \sum_{i=1}^{N_F}
        \frac{\partial^2}{\partial m_i \partial (1/T)} \ln Z
\label{suscep_therm}
\eqn
Assuming that the free energy is dominated by its
singular part, Eq.~(\ref{free_energy}) then leads
to the scaling predictions for the peak heights
of the susceptibilities at the line of pseudo-critical
couplings
\begin{eqnarray}
\chi_m^{\rm peak} & \sim & m^{-z_m} \label{chimpeak} \nonumber \\
\chi_t^{\rm peak} & \sim & m^{-z_t} \label{chitpeak}
\label{scaling}
\end{eqnarray}
where the exponents are given by
$z_m = 2 - 1/y_h$ and $z_t = (y_t-1)/y_h + 1$.
The pseudo-critical line itself is expected to
follow
\beqn
\frac{6}{g_c^2(m)} = \frac{6}{g_c^2(0)} + c m^{z_g}
\label{pseudoline}
\eqn
with $z_g = y_t/y_h$. The values of these exponents
for various symmetries \cite{expo}
are given in table~\ref{tab:exponents}.
At finite lattice spacing the exact chiral symmetry
of the staggered fermion action is $U(1) \simeq O(2)$.
However, sufficiently close to
the continuum limit one expects $O(4)$ exponents.
The possibility of mean-field (MF) exponents arbitrarily
close to the transition has been raised by \cite{KogutKocic}.

\begin{table}
\caption{Critical exponents for $O(2),O(4)$ and mean field (MF).
         The 2 flavor QCD results are given separately for each spatial
         lattice size, with upper values denoting the JLQCD and
         the lower ones the Bielefeld group numbers.}
\begin{center}
\begin{tabular}{@{}l@{\hspace{3mm}}
                 c@{\hspace{2mm}}c@{\hspace{2mm}}c @{\hspace{3mm}}
                 l@{\hspace{1mm}}l@{\hspace{1mm}}l@{}}
\hline
      & O(2) & O(4) & MF  & L=8      & L=12     & L=16     \\
\hline
$z_g$ & 0.60 & 0.54 & 2/3 & 0.70(11) & 0.74(6)  & 0.64(5)  \\
      &      &      &     &
        \multicolumn{3}{c}{~\hrulefill~~0.63(6) \hrulefill~} \\
$z_m$ & 0.79 & 0.79 & 2/3 & 0.70(4)  & 0.99(8)  & 1.03(9)  \\
      &      &      &     & 0.84(5)  & 1.06(7)  & 0.93(8)  \\
$z_t$ & 0.39 & 0.34 & 1/3 & 0.47(5)  & 0.81(9)  & 0.83(12) \\
      &      &      &     & 0.63(7)  & 0.94(12) & 0.85(12) \\
\hline
\end{tabular}
\end{center}
\label{tab:exponents}
\end{table}

\begin{figure}[htb]
 \epsfig{bbllx=115,bblly=110,bburx=660,bbury=505,
         file=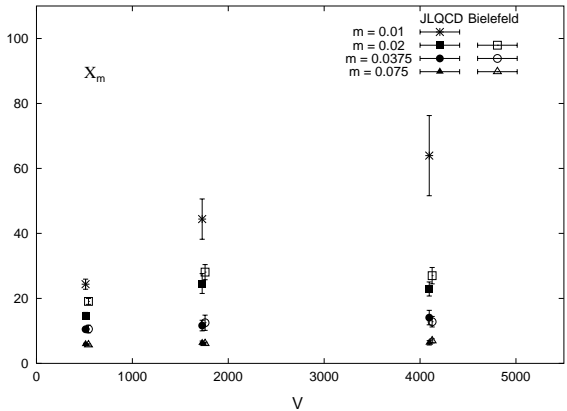,width=75mm,angle=-90}
\vspace{-30mm}
\caption{Volume dependence of the chiral susceptibility at
peak.  }
\label{fig:L_dep}
\end{figure}

Earlier investigations of the exponents on small lattices
($8^3 \times 4$)
had observed partial agreement with $O(4)$ scaling \cite{FKEL}. 
These studies have been repeated
on larger spatial volumes, $L=12,16$. 
In addition to the quark mass values 0.02, 0.0375 and 0.075
JLQCD also ran at $m=0.01$.
The volume dependence of the
chiral peak susceptibility is shown in Figure~\ref{fig:L_dep},
similar results are available for the other quantities.
For $m \geq 0.02$, the susceptibility rises when the volume
is increased from $8^3$ to $12^3$, but then stays 
approximately constant. Thus, a phase transition does not
occur in this mass range, in agreement with earlier claims
\cite{early_crit}. 
At $m=0.01$ the linear increase in the peak height
as the volume is enlarged continues up to $L=16$. As such,
this observation could suggest a first order transition. 
JLQCD however have studied the volume dependence of 
a double-peak structure
in the distribution of the chiral order parameter and conclude
that a first order transition is likely to be absent \cite{JLQCDcrit}.

\begin{figure}[htb]
 \epsfig{bbllx=115,bblly=105,bburx=670,bbury=505,
         file=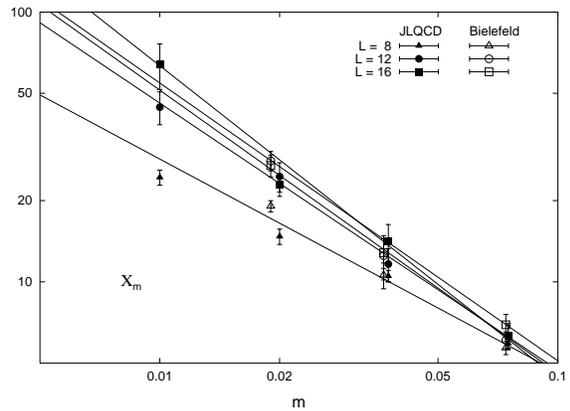,width=75mm,angle=-90}
\vspace{-30mm}
\caption{Mass dependence of the chiral susceptibility at
peak. The upper 4 lines are fit results with Eq.~\refe{scaling}
to the $L=12$ and $16$ data while the lowest line shows the
slope of $O(4)$ scaling.}
\label{fig:m_dep}
\end{figure}

The quark mass dependence of $\chi_m^{\rm peak}$ is shown in 
Figure~\ref{fig:m_dep}, together with fits to the
expected scaling behavior, Eq.~\refe{scaling}.
The resulting values for the critical exponents are
also summarized in Table~\ref{tab:exponents}.
For $z_g$, within two standard deviations agreement with
all three predictions is obtained. For the other two exponents,
both groups consistently observe a drastic change when the
volume is increased from $L=8$ to $L=12,16$. While for
the small volume the value for $z_m$ is in rough agreement
with $O(2)$ and $O(4)$, the results from $L=12$ and $16$
do not agree with any of the predicted numbers.
Indeed, the observed value $z_m \simeq 1$ would be expected
for a first order transition. The thermal exponent $z_t$
is larger than any of the predictions for all volumes.

\begin{figure}[t]
\hspace{-5mm}
\epsfig{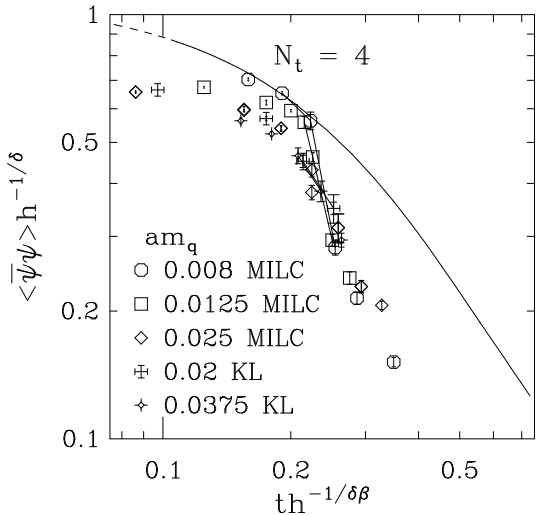}
\vspace{-10mm}
\caption{The magnetic equation of state, Eq.~\refe{mag_eos},
at $N_{\tau}=4$ \cite{Carleton}. The data
at quark masses 0.0375, 0.02 and 0.025 comes from $L=8$ lattices
while at 0.0125 and 0.008 lattice extents up to $L=24$ have been
used. The line is the $O(4)$ scaling prediction. It can be moved
horizontally as well as vertically by adjusting two free normalization
constants.} 
\label{fig:MILC}
\end{figure}

Another way to study the scaling behavior is to compute
the (magnetic) equation of state \cite{Carleton94} 
\beqn
\langle \overline{\psi} \psi \rangle h^{-1/\delta}
= \phi( t h^{-1/\beta\delta})
\label{mag_eos}
\eqn
and compare it with the scaling function $\phi$ 
as determined from a parametrization of $O(4)$ simulation
results \cite{Toussaint}, after adjusting two non-universal
normalization constants.
This has been done by the MILC Collaboration \cite{Carleton}
for $N_{\tau} = 4, 6,8$ and 12. The result for $N_{\tau}=4$ is
shown in Figure~\ref{fig:MILC}. While the data for the larger
quark masses and smaller volumes are compatible with $O(4)$,
the new data at smaller quark masses and larger lattice extent
again show drastic disagreement. (MILC also has looked at one of the
exponents, with the result $z_t = 1.3(2)$.) When $N_{\tau}$
is increased, thus going to smaller lattice spacings, the agreement
becomes increasingly better \cite{Carleton}, 
but it should be remarked that the data
at $N_{\tau}=12$ originates from quark mass values 
$m/T \simeq 0.1$ which are of about the same size as the
larger quark masses used at $N_{\tau}=4$. Also, even at large
spacing one would expect $O(2)$ behavior which is indistinguishable
from $O(4)$ with the current precision of the data.

At the moment there is no convincing explanation for these
discrepancies at hand. In view of the results presented in section 3.3
obtained with an improved gauge action and Wilson fermions
one might speculate that at strong coupling and for the
standard action the relation Eq.~\refe{eq:def} between the QCD parameters
and the thermodynamic variables as they enter the singular part
of the free energy is strongly distorted. More studies
at weaker coupling or with improved actions would be needed
to solve this important question.


\subsection{$U_A(1)$ restoration}

\begin{figure}[t]
 \epsfig{bbllx=10,bblly=120,bburx=531,bbury=650,
         file=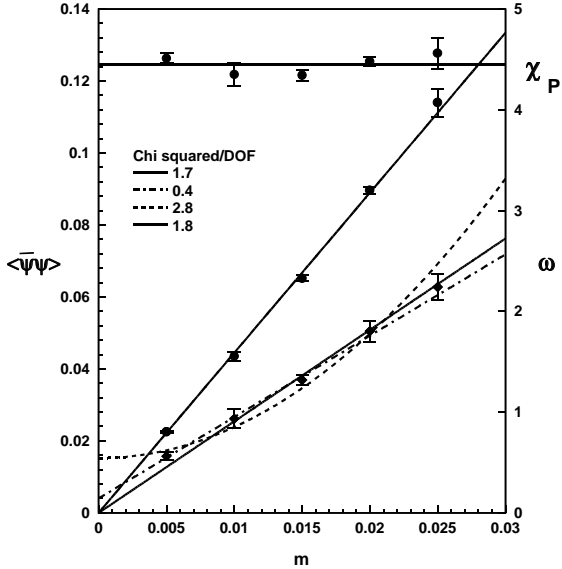,width=75mm}
\vspace{-10mm}
\caption{The quantity $\omega$, Eq.~\refe{ua1}, which measures
the breaking of the $U_A(1)$ symmetry plotted versus the quark 
mass (lowest data) together with various fits \cite{Christ}. 
Also shown are the results
for the integrated pion correlator, $\chi_P$, and the chiral
condensate. The data were obtained on a $16^3\times 4$ lattice
at fixed $\beta$ slightly above $\beta_c$.}
\label{fig:ua1}
\end{figure}

The nature of the chiral transition for two flavors is
strongly affected by the realization of the 
$U_A(1)$ symmetry \cite{WiPi}.
At very high temperatures topologically non-trivial
configurations are suppressed, thus leading to the
effective restoration of the 
symmetry despite the anomaly. For 2 light quark flavors
the effective restoration of $U_A(1)$ is reflected in
the degeneracy of the pion and the isovector-scalar
$a_0 (\delta)$ mass \cite{Shuryak}. This degeneracy
can also be detected by comparing the correlation functions
\beqn
\omega = \int d^4 x ( \langle \pi(x) \pi(0) \rangle
- \langle a_0(x) a_0(0) \rangle )
\label{ua1}
\eqn
If $U_A(1)$ is restored this quantity should vanish in the
chiral limit.
The (generalized) susceptibility has been looked at
by various groups \cite{MILC_ua1,Christ,us_crit}. 
At finite quark mass $\omega$ is dropping
across the chiral transition, but stays non-zero
above $T_c$. Figure~\ref{fig:ua1} shows the latest Columbia
data for this quantity, together with various extrapolations
to the chiral limit.

In the continuum, the susceptibility $\omega$ is expected to be
an analytic and, for $N_F=2$, even function in the quark 
mass. Indeed,
fits with a quadratic $m$ dependence work and lead to a
finite intercept in the chiral limit. However, the data
look strikingly linear and fitting them with a linear ansatz
results in a vanishing of the susceptibility at $m=0$.
At finite lattice spacing, due to zero-mode shifts and
taking the square root of the determinant the approach
towards the chiral limit is not so clear \cite{Christ}.
Therefore one should continue to study the quark mass
dependence at even smaller quark masses as well as at
smaller lattice spacings.

The approach chosen in
\cite{HT_ua1,Lagae} is to determine screening masses.
Above the critical temperature, the difference
between $\pi$ and $a_0$ mass drops considerably, but
a non-degeneracy remains at finite quark mass, thus
confirming the findings originating from the analysis
of the susceptibility. In order to address the problem
of the chiral limit from a different angle,
ref.~\cite{Lagae} also computed the lowest eigenvalues 
$\lambda$ and
corresponding eigenvectors $\psi_{\lambda}$ of the
fermion matrix. In the continuum, in the phase symmetric
with respect to the axial $SU(2)$,
the chiral limit of $\omega$ is given by the zero-modes,
\beqn
\omega = \langle \sum_{\lambda=0} \frac{{\overline
\psi}_{\lambda} \gamma_5 \psi_{\lambda}}{i \lambda + m} \rangle
\label{zeromode}
\eqn
In \cite{Lagae} it is then verified that $\omega$ 
obtained in the standard fashion is saturated by the
contribution from low eigenmodes thus supporting
that continuum arguments seem to hold at lattices spacings
of ${\cal O}(0.15 {\rm fm})$ ($N_{\tau}=8$). The contribution
of zero modes in Eq.~\refe{zeromode} is to leading order
quark mass independent. If this could be verified e.g. by running
at more quark mass values the results of \cite{Lagae} would
indicate that the $U_A(1)$ symmetry is not restored at the
chiral transition.

\vfill
\newpage

\subsection{Wilson fermions $N_F=2$}

The phase diagram of QCD at finite temperature with 2 flavors
of Wilson quarks has been clarified in \cite{Aoki} and
has been explained in great detail in last year's review \cite{Akira96}.
At finite $N_{\tau}$,
the line $\kappa_c(\beta)$ defined through the vanishing of
the pion mass starts off at 1/4 at $\beta=0$ and extends
to $\kappa_c \simeq 0.22$ at about $\beta \simeq 4.0$
where it bends backwards again to the region
of stronger couplings 
(see also \cite{Bitar_97}).
On the other hand, coming from the confined phase,
at the thermal line $\kappa_t(\beta)$ where the Polyakov loop
develops a non-vanishing expectation value the pion mass
increases rapidly due to the approximate restoration of chiral
symmetry. Only in the region where the thermal line 
is close to $\kappa_c$ does the theory have a pion with a small mass.
Thus, the chiral transition can only be explored in that region.
Unfortunately, this region is at strong coupling for $N_{\tau}=4$
and moves towards smaller coupling only very slowly 
with increasing temporal extent of the
lattice \cite{Iwa_kappac}, rendering a study of the transition
in the vicinity of continuum physics prohibitively expensive.
For that reason and for the well-known pathologies \cite{Bernard3}
several groups have started to work with improved actions.

\begin{figure}[htb]
 \epsfig{bbllx=70,bblly=110,bburx=515,bbury=635,
         file=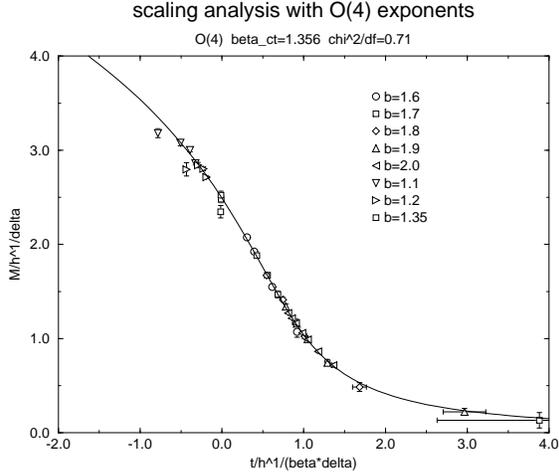,width=63mm,angle=-90}
\vspace{-10mm}
\caption{The magnetic equation of state, 
Eq.~\refe{mag_eos}, 
with two flavors of standard Wilson fermions 
on improved glue \cite{Kaya}.
}
\label{fig:wilson_O4}
\end{figure}

Iwasaki et al. simulated with the standard Wilson fermion action
but on RG-improved glue \cite{Kaya}. Qualitatively, the phase
diagram is very similar to the standard one so that small pion
masses again are obtained in the vicinity of the finite
temperature $\kappa_c$ cusp. In addition to the phase diagram
the group has also investigated
the magnetic equation of state, Eq.~\refe{mag_eos}. 
For Wilson fermions quark mass and chiral order parameter
have to be obtained from chiral Ward identities \cite{Boccia}.
This involves renormalization constants for which the lowest
order perturbative values have been used in \cite{Kaya}.
The results, including new
data at low $\beta$ are shown in Figure~\ref{fig:wilson_O4}. 
The agreement with the $O(4)$ scaling curve is remarkable.
The analysis was carried out on 
lattices of size $8^3 \times 4$ and 
with mainly not very small quark masses. It would be very
interesting to continue the investigation on larger lattices
and with more data at smaller quark masses.

The MILC Collaboration has analyzed the finite temperature
transition with the Symanzik-L\"uscher-Weisz gluon 
action in combination with
the Sheikholeslami-Wohlert fermion action, both tadpole improved
\cite{MILC9703}.
They obtain a much smoother cross-over behavior as
compared with the standard actions which suggests that the pathologies
of the latter are lattice artefacts. The results for meson
masses and the string tension have been presented already
in section 2.
Also Bielefeld started to work with Wilson fermions.
Preliminary results on the phase diagram for tree level-improved
glue and the clover action were presented at this conference
\cite{Oevers}. Qualitatively, the behavior of the $kappa_c$ and
$kappa_t$ lines is very similar to the standard one.

\subsection{Wilson fermions  $N_F \neq 2$}

The phase diagram has also been studied for the number of flavors
differing from 2. All analyses have been carried out with
standard Wilson fermions on lattices with $N_{\tau}=4$ so far. 

The case $N_F = 1$ has been looked at in
\cite{Dina}. For heavy quarks the deconfinement transition 
shows first order behavior as a finite size analysis of the
Polyakov loop reveals. The first-order transition weakens
when the quark mass is decreased. The
end-point of the transition is estimated to occur at about 1.1 GeV.

The cases $N_F = 3$ and 4 have been studied in Tsukuba 
\cite{nfvier}. In both cases the phase diagram is very
similar to the one with 2 flavors, in particular the 
$\kappa_c$ line forms a cusp. 
Regarding the nature of the chiral transition, for
$N_F \geq 3$ one expects first order in the
continuum limit \cite{WiPi}.
At large quark masses, away from the cusp, one indeed
observes first order behavior. When the quark mass is
lowered however, the $N_F=4$ data shows a weakening
of first-order signals. For $N_F=3$, one has to use
an approximate algorithm e.g. the Hybrid R. At 
a time step size of $\delta \tau = 0.01$ two-state signals
were observed also close to $\kappa_c$. These signals
weaken when the time discretization errors of the algorithm
are decreased ($\delta \tau = 0.005$). Thus, for both
$N_F=3$ and 4 the order of the transition is still unclear.

\section{EQUATION OF STATE}

\begin{figure}[t]
 \epsfig{file=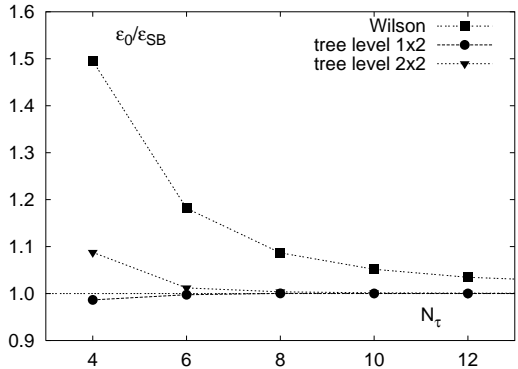,width=75mm}
\vspace{-15mm}
\caption{The gluonic part of the energy density at
high temperature computed on lattices with finite
temporal extent in units of
the continuum Stefan-Boltzmann value
for various gauge actions \cite{FKtsu}.}
\label{fig:eos_anal_pg}
\end{figure}

\begin{figure}[htb]
 \epsfig{file=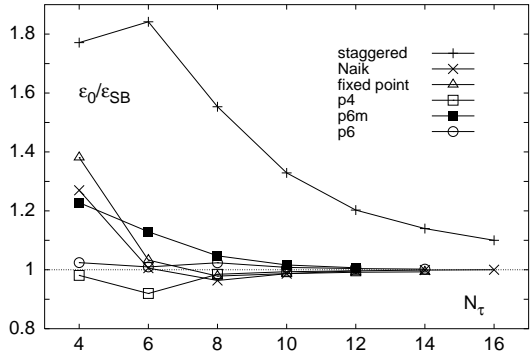,width=75mm}
\vspace{-15mm}
\caption{The fermionic part of the energy density at
high temperature computed on lattices with finite
temporal extent in units of
the continuum Stefan-Boltzmann value
for various staggered fermion actions \cite{Peikert}.}
\label{fig:eos_anal_f}
\end{figure}

Thermodynamic quantities like pressure or energy density
receive substantial contributions from high momentum
modes, $p = \pi T$. Since on the lattice 
these are distorted by UV
cut-off effects e.g. the energy density in the infinite
temperature limit deviates considerably from the continuum
Stefan-Boltzmann value. For the pure gauge theory 
in the standard Wilson discretization the corrections are 
\begin{eqnarray}
\lefteqn{\epsilon_0^G = \epsilon_{SB}^G \left[ 1 
+ \frac{10}{21} \left( \frac{\pi}{N_{\tau}}\right)^2 \right.}
~~~~~~~~~ \nonumber \\
& & \left.+ \frac{2}{5} \left( \frac{\pi}{N_{\tau}}\right)^4
+{\cal O}( \left( \frac{\pi}{N_{\tau}}\right)^6) \right]
\label{eos_pg}
\end{eqnarray}
This effect is even larger for the fermionic part of the
energy with standard staggered quarks,
\beqn
\epsilon_0^F = \epsilon_{SB}^F \left[ 1 
+ \frac{465}{441} \left( \frac{\pi}{N_{\tau}}\right)^2
+{\cal O}( \left( \frac{\pi}{N_{\tau}}\right)^4) \right]
\label{eos_f}
\eqn
On the other hand, the signal for this quantity
vanishes proportional to $1/N_{\tau}^4$. 
Bulk thermodynamic quantities were investigated in the
standard discretization, both quenched \cite{BoydEOS} 
and with 2 flavors of staggered quarks \cite{MILCEOS}.
While for the quenched theory it was possible to
extrapolate to the continuum, in view of the
analytic results, 
Figures~\ref{fig:eos_anal_pg},\ref{fig:eos_anal_f},
this seems to be very
difficult with dynamical fermions.


In the infinite temperature, Stefan-Boltzmann limit,
cut-off effects are reduced at tree level. 
Perturbative or tadpole improvements do not have any effect
in this limit.
The size of the remaining corrections to the Stefan-Boltzmann limit
can be read off 
Figures~\ref{fig:eos_anal_pg},\ref{fig:eos_anal_f}
where a variety of different actions are compared
with each other \cite{FKtsu,Peikert}. 

The effect of various improvement schemes at temperatures
as low as the critical temperature has been tested in
quenched simulations \cite{Beinlich_97}. For the pressure,
it seems that tree level improvement is the leading effect,
although one would have expected that close to the transition
infra-red modes and their improvement would be more important.
Tadpole improvement seems to have an effect though for
the interface tension \cite{BKP}.

A first attempt to analyze bulk thermodynamic quantities
by means of an improved fermion discretization scheme,
the Naik action, has been published recently \cite{impferm}.
At the moment, there are investigations underway which
try to estimate the effect of various improvement strategies
on the restoration of flavor symmetry 
\cite{Luo,Sinclair,MILC_fat,Peikert}. 
It remains to be seen how much this would
help to extract energy density or pressure closer to
the continuum limit at finite temperatures.

\section{SCREENING LENGTHS AND MASSES}

The study of hadronic excitations at finite temperature
is an important subject as it reveals more details about
the properties of the plasma phase as well as the approach
to the transition where temperature dependencies could
mimick plasma signatures.

The temperature dependence of pseudoscalar and vector meson
(screening) masses has been studied in \cite{Peter}.
The masses were determined at temperatures slightly below and
above the transition, $T \simeq 0.9 T_c$ and $T \simeq 1.2 T_c$
respectively, and compared with zero temperature results. The
investigation was carried out with the Sheikholeslami-Wohlert
fermion action applied to quenched gauge configurations. The
results, obtained in the quark mass region ranging from
light to approximately the charm quark masses do not exhibit 
any strong temperature effect for $0.9 T_c$. At $1.2 T_c$
the data are approaching the high temperature, free quark limit,
$M_{\pi,\rho} \lsim 2 \pi T$. This confirms earlier results
for staggered fermions \cite{FK93}. 

Heavy quarks at temperatures around the critical one are investigated 
in \cite{Jochen}. The heavy quarks in the range
$m_c < m < m_b$ are simulated by means of an
NRQCD action applied to quenched configurations 
which have been generated
with a $(1 \times 2)$ tree level improved gauge action.
The propagation of quarkonia states is followed in the
time direction. For that purpose,
the investigation is carried out on anisotropic lattices
with a large anisotropy ratio
$\xi = a_{\sigma} / a_{\tau} = 4.65$ in order to have
enough Matsubara frequencies. So far, the analysis
was done for the $^3 S _1$ ground and first excited state.
Below $T_c$, at about $0.8 T_c$ no temperature effect was
seen. At $1.2 T_c$, the propagator which is dominated by
the ground state at large time separations $t$ from the source 
becomes flatter than the
zero temperature propagator at large $t$. 
This effect can be interpreted as a decrease in the mass
by a small amount of about 12 MeV at the lightest quark mass
simulated. The effect becomes weaker with increasing quark mass.
The first excited state,
projected on by the same trial wave function as at $T=0$,
undergoes a larger change of about - 240 MeV at the charm quark mass.
At least qualitatively, the results are in accord with the expectations
from a Debye-screened potential model, namely that smaller states feel
the screening less.

Investigations of the Debye mass itself have been presented by 
\cite{Kari,Rank}. The aim of both groups is to confront the
lattice data with perturbative results. It is known since
long \cite{Linde} that high temperature QCD is plagued by
infrared divergencies. These are cured at ${\cal O}(gT)$
by the Debye or electric gluon mass $m_e$. It is widely believed
that the magnetic mass
$m_m$ is non-vanishing and acquires a value of ${\cal O}(g^2T)$.
Such a mass would solve
the infrared problem at this order but cannot be calculated in
perturbation theory. If $m_m$ is non-vanishing 
next-to-leading order perturbation theory predicts \cite{Rebhan}
for the electric mass
\begin{eqnarray}
\lefteqn{m_e^2 = m_{e0}^2 ~ \left(~1~+ \right.}  
~~~~~~~~~~~ \nonumber \\ 
& & \left.g \frac{N}{2 \pi} \sqrt{
\frac{6}{2N+N_F}} [\ln\frac{2m_e}{m_m} - \frac{1}{2}]\right)
\label{debye_pert}
\end{eqnarray}
where $m_{e0}$ denotes the leading term
$m_{e0} = \sqrt{N/3+N_F/6} gT$.
The methods used by \cite{Kari} and \cite{Rank} are quite different.
In \cite{Kari} dimensional reduction is applied and the 3d
effective theory is simulated. The Debye mass is then extracted from the
correlation of a gauge-invariant operator \cite{Arnold}.
The authors of \cite{Rank} generate $SU(2)$ gauge configurations with the
standard as well as an improved action. They analyze gauge-variant
gluon and Polyakov loop correlations after fixing to Landau gauge.
Their data show a non-vanishing magnetic gluon mass which
varies with temperature according to $m_m(T) = c g^2(T) T$.
For the electric mass, the results of the
two groups differ somewhat quantitatively. However, both groups 
show clearly that
(next-to-leading order) perturbation theory describes
the data only at very large temperatures of ${\cal O}(10^4 T_c)$.

\section{QCD AT FINITE DENSITY}

It is well known that at finite chemical potential
$\mu \neq 0$ the effective action of QCD becomes
complex. This prohibits the use of standard Monte Carlo
methods to evaluate the path integral. A way to circumvent
these difficulties is provided by what is known as the
Glasgow method \cite{Ian}. This method was applied
to QCD at intermediate and recently at strong coupling
\cite{Maria}.
In the latter case mean-field computations \cite{meanfield} 
and results 
from a monomer-dimer simulation \cite{monomer} are available.
The results of \cite{Maria} show a finite
density transition at a critical value $\mu_c$ which agrees
with the mean-field as well as with the monomer-dimer number.
However, also an onset $\mu_o$ at which the
number density starts to deviate strongly from 0 is observed. 
The value of $\mu_o$ coincides with $m_{\pi}/2$ 
for smaller quark masses so that in the chiral limit 
the chiral condensate would immediately drop from its
zero chemical potential value when $\mu$ is increased
from 0. Similar findings were obtained at intermediate
coupling. The same behavior was also
seen in the quenched approximation where it was traced back
to the failure of this approximation at 
finite $\mu$ \cite{Gocksch}. 
The reason for this behavior in full QCD is not clear.
One might speculate \cite{Ian} that sampling configurations at
$\mu = 0$ might not catch the right physics at finite $\mu$.
Similar observations on $\mu_o$ were reported in \cite{Galante}
where the modulus of the determinant was used in the partition function.
A different approach has been presented in \cite{Susan} where
perturbatively irrelevant four-fermion interactions were added
to the QCD action. This way one is able to simulate at vanishing
bare quark mass, but it has to be clarified yet whether this
approach really produces only physical results.

\section{CONCLUSIONS}

The results of last year's research on finite temperature lattice QCD 
with dynamical quarks have been reviewed.
At currently accessible quark masses the critical temperature
for 2 flavors is being estimated as around 150 MeV.
This number is, not unexpectedly, quite smaller than
the quenched value of 270(5) MeV. Correspondingly, at
fixed $N_{\tau}$ the computations with quarks have been
carried out at considerably larger lattice spacings compared
to quenched simulations. Moreover, going to lighter quarks 
drives one to stronger coupling. One may speculate that
lattice spacing effects are the reason for the very unexpected
behavior at the chiral/deconfinement transition for both
standard staggered and Wilson fermions while improving only the gauge 
part already seems to lead to remarkable agreement with the theoretical
predictions. In addition, the computation of the
energy density of a free fermion gas shows that the discretization
effects can be large for fermions when the standard 
lattice transcriptions are used. Thus it is highly desirable
to find improved actions for the analysis of the
thermodynamics with quarks.

\section*{ACKNOWLEDGEMENTS}

It is a pleasure to thank my colleagues at Bielefeld,
B. Beinlich, A. Bicker, J. Engels, F. Karsch, 
C. Legeland, M. P. Lombardo, M. L\"utgemeier, M. Oevers,
A. Peikert, B. Petersson, J. Rank, K. Rummukainen and P. Schmidt
as well as
S. Aoki, I. Barbour, K. Bitar, G. Boyd, N. Christ, C. DeTar, 
J. Fingberg, Y. Iwasaki, S. Kaya, J.-F. Laga\"e, 
S. Morrison, M. Okawa and A. Ukawa for (preliminary) data,
plots, discussions and sharing their insight.

\end{document}